\documentclass[preprint,12pt,3p]{elsarticle}

\usepackage{times}
\usepackage{graphicx}

\usepackage{amssymb}
\usepackage{amsmath}                     
\usepackage{amsfonts}                    
\usepackage{amsthm}    
\usepackage{xfrac}
\biboptions{round, authoryear}

\renewcommand\cite{\citep}

\journal{Earth and Planetary Science Letters}

\begin{document}

\begin{frontmatter}

\title{A Model of the Primordial Lunar Atmosphere}

\author[label1]{Prabal Saxena\corref{cor1}}
\address[label1]{NASA/Goddard Space Flight Center,\\8800 Greenbelt Rd, Greenbelt, MD 20771, USA}

\cortext[cor1]{Corresponding author}

\ead{prabal.saxena@nasa.gov}

\author[label2]{Lindy Elkins-Tanton}
\address[label2]{School Of Earth \& Space Exploration, Arizona State University\\PO Box 876004, Tempe, AZ 85287, USA}

\author[label1]{Noah Petro}

\author[label1]{Avi Mandell}

\begin{abstract}
We create the first quantitative model for the early lunar atmosphere, coupled with a magma ocean crystallization model.  Immediately after formation, the moon's surface was subject to a radiative environment that included contributions from the early Sun, a post-impact Earth that radiated like a mid-type M dwarf star, and a cooling global magma ocean.  This radiative environment resulted in a largely Earth-side atmosphere on the Moon, ranging from $\sim$10$^4$ to $\sim$10$^2$ pascals, composed of heavy volatiles (Na and SiO). This atmosphere persisted through lid formation and was additionally characterized by supersonic winds that transported significant quantities of moderate volatiles and likely generated magma ocean waves.  The existence of this atmosphere may have influenced the distribution of some moderate volatiles and created temperature asymmetries which influenced ocean flow and cooling.  Such asymmetries may characterize young, tidally locked rocky bodies with global magma oceans and subject to intense irradiation. 
\end{abstract}


\end{frontmatter}


\section{Introduction}
\label{sec1}
The Moon is an important and relatively observationally accessible marker of solar system history.  Evidence obtained from study of the lunar surface has yielded insight into numerous physical processes that have transformed the planets over the history of the solar system \cite{1991lsug.book.....H, 1997GeCoA..61.2331K, 2001SSRv...96....9S}. Consequently, interpretation of the evolution of the lunar interior and surface helps to inform understanding of the environment the Earth evolved in. Critical to the interpretation is tracing the history of the surface right from the initial formation of the Moon to the present day.

Petrological evidence supports the theory that the early lunar surface was very different from the solid, crater marked version observed today.  Instead,  the Moon was believed to have been covered by a deep global magma ocean immediately after its formation \cite{GRL:GRL29302}.  This magma ocean is believed to be a consequence of the relatively quick accretion of the Moon after its formation from a collision between the Earth and a planetary sized impactor \cite{2004Icar..168..433C,2016arXiv160808959B}. After formation, the magma ocean cooled and solidified to $\sim$70-80$\%$ in less than 1000 years, leading to the eventual flotation of plagioclase rockbergs \cite{1977LPSC....8..601L} that would form a floatation lid. 
The initial radiative environment of the Moon's surface was likely to have an additional consequence often not associated with the Moon - the creation of an early atmosphere.

While today the Moon possesses a rarefied exosphere composed largely of Argon, Helium and Neon \cite{2016LPICo1911.6022E}, previous work has recognized that at earlier times, the Moon likely possessed denser atmospheres \cite{ROG:ROG1541}. 
These atmospheres are proposed to have been a result of the vapor pressure equilibrium that likely existed above the exposed magma on the surface.   However, several studies have also recognized that an immediately post-collision Earth may have been hot enough to induce a temperature asymmetry on the Moon \cite{WASSON1980752,ROG:ROG1541, 2014ApJ...788L..42R}, and that the radiative contribution from the glowing Earth may have served as an additional energy source for vaporization of surface magma.  This additional radiation source and the potential that it may have induced asymmetries on the early Moon may have had some important consequences for surface evolution. 

However, to date there is no model which looks at spatially resolved atmospheric-surface conditions on the Moon prior to and during plagioclase lid formation. Here we discuss results of the first spatially resolved surface model for the early Moon, one that couples an atmosphere model to a Lunar Magma Ocean (LMO) crystallization model.  We use an atmosphere model originally developed to explore the meteorology of Io \cite{1985Icar...64..375I} in conjunction with a LMO crystallization model that yields crystallization timescales for the LMO and surface temperatures as the magma evolves \cite{2011E&PSL.304..326E}.  Our model includes radiative contributions from a cooling lunar magma ocean, the Earth immediately after the Moon forming impact and the early Sun.  The following two sections describe the details of the atmosphere and magma ocean crystallization models and how they were linked.  The results of the model are then given in the next section. Finally, we conclude with a section discussing potential implications of the results and consideration of alternate Moon formation and evolution models.

\section{Atmospheric Model Details}
\label{sec2}

Our atmosphere model is a one dimensional vertically integrated model, solving a system of equations for conservation of mass, momentum and energy.  The structure of our model is the atmospheric model described in \citet{1985Icar...64..375I}.  This model solves for the atmospheric pressure, temperature and velocity (at the base of the atmosphere) as a function of the angular distance away from the substellar point (the form of the conservation equations used are given in \citet{1985Icar...64..375I}).  The equations used for the model are given in the Supplementary Material.

Values in the equations for the mass per atom, m, as well as C$_{p}$ are taken for the expected dominant constituent in the atmosphere (which in most cases is sodium). The choice for a single constituent atmospheric model is based upon expected vaporization pressures for a Bulk Silicate Earth (BSE) \cite{2011ApJ...742L..19M}.  The choice of a BSE composition versus a Lunar Primitive Upper Mantle (LPUM) \cite{2006GeCoA..70.5919L} or Taylor Whole Moon (TWM) \cite{1982pslp.book.....T} composition was made in order to remain agnostic about potential mechanisms for moderate volatile loss - in particular to avoid the assumption that all or most of the apparent moderate volatile depletion occurred during formation (particularly given evidence of potential increased CME activity and incidence early in the Sun's history).  These vaporization pressures are calculated using the MAGMA code \cite{1987E&PSL..82..207F,2004Icar..169..216S}, which calculates the equilibrium between the melt and vapor in a magma exposed at temperatures higher than 1000 K for Al, Ca, Fe, K, Mg, Na, O, Si, Ti and their compounds.  Vaporization pressures as a function of temperature were fit to the Clausius Clapyeron form.  For equilibrium temperatures up to nearly 3500K, the vapor pressure of the dominant constituent is nearly an order of magnitude greater than the next most significant constituent.  This justifies our assumption of a single species atmosphere.  

These vapor pressure curves can be used to extract the constants used in our vapor pressure equations.  Since sodium was the dominant constituent for most models, we explicitly state any models where there was a different dominant atmospheric constituent.  The only case where multiple components were summed was at the point immediately after formation, when SiO was a major component.  In this case we summed the partial pressures to find an overall pressures and restricted motion to the slower of the two velocity profiles.  To determine dynamic viscosity we used Sutherland's formula.  We use the values (see Supplemental Information) listed in \cite{2011ApJ...743L..36C} for the equation (it is important to note that Sutherland's formula is only valid to about 555K, but simulations we ran show that our results are not highly sensitive to small extrapolated temperature appropriate variations).  

The radiative environment of the early Moon controls the surface temperature for the atmospheric model.  Inputs for the surface temperature included the radiative contribution of the early Sun, the Earth immediately post-Moon formation impact, and the surface temperature of the Lunar Magma Ocean.  The farside temperature and spatially uniform contribution of the Sun’s radiation we used corresponded to a solar flux $\sim$70\% of the present day value.  
The spatially uniform contribution of solar flux is a simplification since the rotation of the Moon would lead to a diurnal cycle.   However, given the short rotation timescales for a tidally locked early Moon ($\sim$ 0.3 - 0.75 Earth days) and the relatively small radiative contribution of the Sun compared to the Earth at the Moon (about an order of magnitude less), such an approximation is a reasonable first order simplification.  Given the relative magnitudes of the two fluxes, a diurnally varying Solar flux is unlikely to change the overall atmospheric profiles significantly.  It would most likely create a time varying asymmetry in the extent of the atmosphere and wind magnitudes on the two sides of the sub-Earth point.

The contribution of the radiation due to a hot Early Earth is obtained by taking radiating temperature values given in \cite{Zahnle2007, Zahnle201574}.  Moon formation simulations that indicate high outer layer temperatures for the Earth after the collision underpin the prediction of high radiating temperatures for the Earth used in this study.  The steep drop in radiating temperature, particularly as the Earth may develop a steam atmosphere, occurs after the time period corresponding to lid formation on the Moon.  Earth radiating temperatures (surface temperatures are much higher) used as inputs for the three times the models were output for were 2500, 2450 and 2300 K. 

Radiative input from Earth was attenuated as a function of angle of incidence by including a disk approximated angular size of the Earth as observed on the lunar surface.  We model the Earth's radiative contribution to the Moon using a lambertian profile used in \cite{2011ApJ...743L..36C} (with a sub-Earth temperature calculated for an albedo of 0.3, which we consider conservative given the low albedo of the similarly hot 55 Cnc e \cite{2016Natur.532..207D}) and extending it to the total Earth illuminated portion of the Moon, which is limited by the effective angular size of the Earth in the Moon's sky.  This is done by using this temperature profile for 0 $\leq$ $\theta$ $\leq$ 90 and mapping that temperature profile to 0 $\leq$ $\theta$ $\leq$ 90 + 0.5$\theta_{*}$, where $\theta_{*}$ is the approximate angular size of the Earth as seen from the sub-Earth point of the Moon and is given by $\theta_{*}$ = 2 arctan (R$_{*}$/a), where a is the Moon’s orbital distance.  The orbital separation of the early Moon (which is expected to be tidally locked  $\leq$  $\sim$100 days) from the Earth is derived from equation 1 of \cite{WASSON1980752} but cases are also tested for slower and faster migration (with similar overall results - see \ref{Alternate}).

Intuitively, this roughly takes into account the penumbra effect of illumination due to the angular size of the Earth by treating the Earth as a continuum of point sources that consequently illuminate slightly shifted portions of the Earth-side.  This is an approximation as it ignores the overlap of illumination between those adjacent points, but it still provides a very similar temperature model to those used in analogous work \cite{2011ApJ...743L..36C, 2011Icar..213....1L} (differences in the illuminated portion of the tests we ran for planets used in those studies are less than half a degree).  

Finally, the last input for surface temperature is the top of the LMO temperature.  This temperature is conservatively assumed to be liquidus for the evolving magma (ignoring contributions such as radiogenic heating).  We use the magma crystallization model in the following section to model the evolution of the magma and the consequent top of the LMO temperature.  A rough estimate of the total net heat loss from the Moon over time can be approximated using the change in temperature of the evolving magma summed with the latent heat lost due to crystallization.  

There are several details which are not considered in our atmosphere-magma ocean model.  All of these have been neglected due to what is either their relatively minimal effect on the bulk surface properties or in order to remain as conservative as possible regarding the radiative inputs to the atmosphere.  The atmosphere model neglects the effect of rotation as a first approximation. While Rossby numbers are larger but on the order of unity and rotation may be useful to model in the future, rotation terms are unlikely to effect the qualitative results of the model as it pertains to this study. We do not include absorption or scattering effects of the two atmospheric constituents in the model, Na and SiO.  Atomic sodium is only somewhat opaque in the visible portion of the spectrum, and is unlikely to have significant influence on radiative transfer in the infrared, where incoming flux from the glowing Earth would peak. SiO, on the other hand, is fairly opaque in the infrared.  However, SiO is a major constituent only for a very short period initially after formation of the Moon.  Its infrared opacity is likely to produce lower atmospheric temperatures.  This would serve as a negative feedback as the lower temperature would condense SiO and leave the Na dominant atmosphere which characterizes most of the existence of the atmosphere.  Qualitatively, this means that our estimate of the 0.001 Kyr pressure, temperature and wind velocity profiles are likely an upper limit given the other inputs we used.

Neither the latent heat due to crystallization in the LMO or due to vaporization/condensation of atmospheric constituents in considered.  The latter terms are likely to reduce temperatures  near the sub-Earth point and raise them near the more tenuous part of the atmosphere near the ‘terminator’.  However, contributions are small enough to be neglected in a first approximation.  On the other hand, latent heat due to LMO crystallization would only add another heat source that would raise surface temperatures through time.  However, given that such a contribution would only extend the atmospheric profiles a little farther in time, such an effect is unlikely to significantly affect the underlying the conclusions of this study or even the likelihood of other atmospheric and temperature asymmetries inducing mechanisms. Indeed, even if a lower albedo was chosen, slightly higher than liquidus top-of-the-LMO temperatures were chosen immediately after Moon formation, and latent heat of crystallization of the LMO was included, it is unlikely there would be significant differences in the effectiveness of different mechanisms.  Potential wind driven or temperature driven advection processes would merely operate more efficiently, while atmospheric temperatures would still be too low over the majority of the Moon immediately after formation for other accretion or atmospheric advection related mechanisms.

\section{Magma Ocean Crystallization Model Details}
\label{sec3}

Our magma ocean crystallization model is the same one described in \cite{2011E&PSL.304..326E} (further detail given in Supplementary Information). That model was also specifically used for the early Moon and consequently details of the set up (including choice of parameters) and implementation of the model can be obtained from that reference.  Additionally, details on the crystallization sequence and the overarching assumptions regarding the ocean are given in the Supplementary Information section.  There are, however, some important changes we made for the purposes of this study.  The first is that we implemented a Bulk Silicate Earth composition for the LMO.  This was easy to implement since assemblages were calculated a priori, and ensured our compositions were consistent throughout our model.  The most often noted differences between a BSE composition versus LPUM or TWM compositions are typically regarding the moderate volatile content of Na and K compounds.  However, these differences are only several factors and since the total mass content of these compounds is very small in all compositions, this does not greatly impact the solidification pathways or heat loss calculations.  Additionally, the total mass of the atmosphere is very small compared to the volatile content (differing by 7-8 magnitudes in even the most volatile poor compositions) even during the periods when the Moon possessed the thickest of its atmospheres.  Given our preference to remain agnostic regarding moderate volatile loss and the similarity in compositions between the Earth and Moon, the choice of an alternative to the BSE composition is unlikely to effect the main findings of this study.

The magma ocean crystallization model was used to determine timescales to lid formation and solidification of the ocean and was also used to determine top of the LMO temperatures for the atmosphere model.  Top of the LMO temperatures were obtained by following an adiabat through to the top of the evolved magma in the same manner as described in \cite{2011E&PSL.304..326E}. Solidification times were also calculated using the same energy balance described in \cite{2011E&PSL.304..326E}, but values for heat loss were adjusted based on the new radiative inputs that were modeled at the surface.  The temperature exterior to the magma ocean was thus higher than the value assumed for a vacuum in the original model and was adjusted over time for different cooling models (described in \ref{CoolSolid}).

\begin{figure}[t!]
 \includegraphics[width=\textwidth]{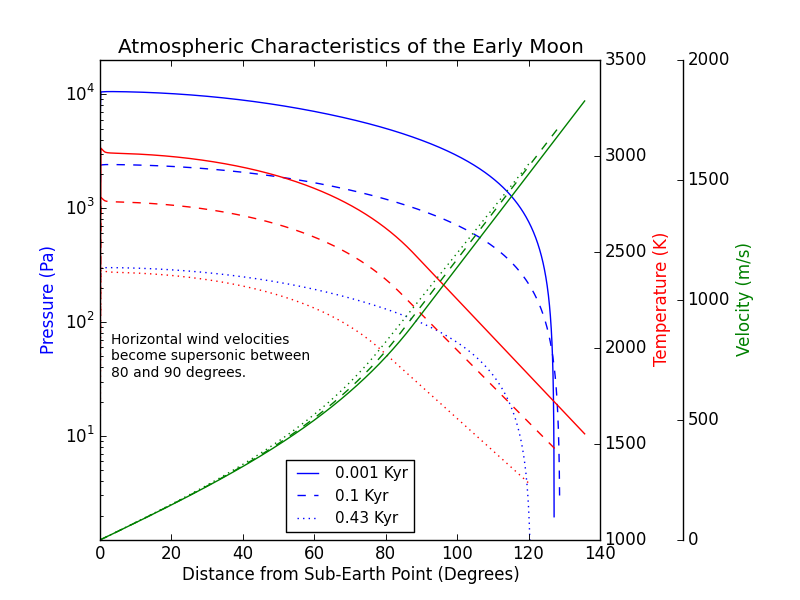}
\caption{Atmospheric profiles for the Early Moon.  Pressure (blue), Temperature (red) and a Horizontal Wind Velocity profile (green) are plotted as a function of angle away from the sub-Earth point.  The times correspond to orbital separations of 3, 4 and 5 Earth Radii.}
\label{fig:atmosprofiles}
\end{figure}

\section{Model Results}
\label{sec4}

The LMO crystallization model was first run to provide an initial set of top of LMO/base of the atmosphere temperatures as a function of time.  These values were fed into the atmosphere model and we then used several different models to determine how long crystallization was delayed by radiative effects of a hot, early Earth (see discussion in \ref{CoolSolid}).  Once new cooling times for crystallization were calculated (including time to lid formation), these were fed back into LMO model in order to produce a more accurate top of the LMO temperature for the atmosphere model.  The duration of atmosphere was linked to lid formation since such an event would greatly limit the volatile reservoir for the collapsing atmosphere.  The existence of an LMO atmosphere and lid formation both occurred within the first 1000 years after formation for the most realistic cooling models.  Model outputs are given for times (1 year, 100 years and 430 years post-tidal locking) that correspond to orbital separations of 3, 4 and 5 Earth Radii.  

Using these inputs from the post-collision Earth-Moon system, we find that the primordial Moon possessed a dynamic and collapsing 10$^4$-10$^2$ Pa metal-dominated atmosphere prior to lid formation.  Immediately after formation, this atmosphere was composed of nearly equal parts Na and SiO$_2$, with SiO$_2$ condensing rapidly as temperatures cooled.  sodium remained the dominant constituent of the atmosphere for the rest of its existence.  A depiction of the atmosphere profiles at different times is given in figure \ref{fig:atmosprofiles}.  

Atmospheres were largely hemispheric with strong pressure and temperature gradients due to the asymmetric radiation from the hot, post-collision Earth.  Immediately after formation, when the Moon was at an orbital distance near the fluid roche limit of the Earth, the lunar atmosphere had a pressure near $10^{4}$ Pa near the sub-Earth point.  Nearly 2/3rds of the moon was enveloped in an atmosphere, with pressures in excess of $10^{3}$ Pa and temperatures greater than 2000K. In the 'dawn' and 'dusk' regions near 110-130 degrees from the sub-Earth point, the atmosphere collapsed.  These strong gradients drove very strong winds directed from the sub-Earth point to the farside.  Horizontal wind velocities became supersonic in the region near 90-95 degrees from the sub-Earth point.  These winds advected substantial energy and mass to the far side.

By the time the Moon reached a distance of 4 Earth radii ($\sim$ 100 years), temperatures were too low to support vaporization of SiO$_2$, and Na was the dominant atmospheric constituent.  Atmospheric profiles of pressure, temperature and horizontal wind velocity remained similar but slightly lower in magnitude compared to the early period.  The spatial extent of the atmosphere also decreased slightly, but the Earth-facing hemisphere still maintained atmospheric pressures of $\sim 10^{3}$ Pa.  Sub-Earth point atmospheric densities were equivalent to approximately mean Mars surface atmospheric density values (densities were almost an order of magnitude higher at 1 year). Horizontal wind velocities remained strong and became supersonic in a the same region as in the prior model.  

Finally, by the time the magma ocean had nearly cooled and crystallized to the point of plagioclase solidification and flotation, the atmosphere had decreased in both pressure and temperature.  By the time the moon had reached an orbital separation of 5 Earth radii, the lunar atmosphere had collapsed to pressures of $\sim 10^{2}-10^{3}$ Pa. Temperatures also decreased to $\sim$ 1500-2300 K, but winds remained very strong and still became supersonic towards the far side.

These winds were also likely to have shaped the evolution of the crust through the generation of waves in the magma ocean.  Wind waves in water are a common feature on Earth and the potential for analogous waves in other media have been examined for bodies such as Early Mars and Titan \cite{Lorenz2005556, 2013Icar..225..403H}.

\section{Discussion}
\label{sec4}

\subsection{Mass Transport: Atmospheric Advection and Magma Ocean Waves}
\label{MassTransport}

Wind wave generation in a liquid body is a long studied problem for the Earth and has been applied to other bodies.  The two main mechanisms we focus on are Capillary-Gravity wave generation by the Miles-Phillips mechanism and classical Kelvin-Helmholtz wind wave generation.  The former has been adapted to study the potential for wave generation on Titan \cite{2013Icar..225..403H} while the latter has been suggested as a more likely mechanism for strong winds between the boundary of two fluids with a large density ratio \cite{ Miles_2006, 2008PhFl...20i4106S}.  Capillary-Gravity waves are the most easily excited on Earth but given the higher viscosities and surface tension of magma, Kelvin-Helmholtz waves may be the dominant wave generation mechanism on the early Moon.  We examine both processes in the context of the Earth-facing hemisphere of the early lunar surface environment, where winds are very strong yet still sub-sonic.

\begin{figure}[t!]
 \includegraphics[width=\textwidth]{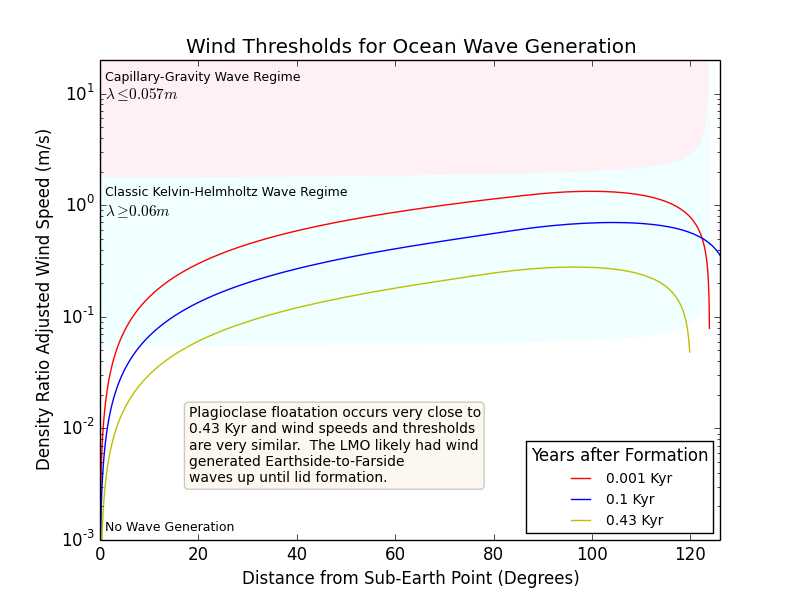}
\caption{Density adjusted wind thresholds for wave generation in the early Moon's magma ocean. Wind speeds were high enough to generate Earth to farside magma ocean waves through nearly the entire existence of the early lunar atmosphere.}
\label{fig:windthres}
\end{figure}

For capillary-gravity wave generation we use the methodology described in \cite{2013Icar..225..403H} to determine the wind threshold needed to excite waves smaller than a critical wavelength.  We use the methodology described in section 2 of \cite{2013Icar..225..403H} which requires solving equations 2.3 - 2.6 in order to determine wind threshold required for wave generation.  One of the strengths of this particular work on capillary-gravity wave generation on Titan is that it considered important differences in kinematic viscosity and surface tension that would arise in different fluids.  We were thus able to use values appropriate for a magma ocean on the early Moon in order to produce results that are likely to be more physically realistic.  In addition to density values for the BSE magma and spatially varying atmospheric density values, we also used a value of  0.375 $N/m$ for the surface tension of the magma \cite{MURASE01111973} and values of 0.01-0.1 $m^{2}/s$ (we use 0.01 for figure 3 and also tested 0.1) for the kinematic viscosity of the magma \cite{2008E&PSL.271..123G, 1999JGR...104.7203J, Lesher2015113, Zahnle201574}.  For Kelvin-Helmholtz wave generation, we used the wind threshold formula given in \cite{Miles_2006, 2008PhFl...20i4106S}, U = $\epsilon^{-1} \sqrt{g/k}$ where $g$ is the gravitaitonal acceleration, $k$ is wavelength and $\epsilon$ is the square root of the ratio of the two fluid densities.  These waves are characterized by wavelengths longer than the capillary length (as indicated by figure \ref{fig:windthres}).   We examine density adjusted wind wave thresholds (adjusting by the value $\epsilon$ in \citet{2008PhFl...20i4106S}), taking into account vertically asymptotic atmospheric wind profiles given in \cite{2013Icar..225..403H} (specifically solving for $U_{\lambda/2}$ in equation 2.6 in that paper using the appropriate density ratios for the fluids and a value of 0.1 $m^{2}/s$ \cite{2008E&PSL.271..123G, 1999JGR...104.7203J} for the magma viscosity).

The results are given in figure \ref{fig:windthres}, which displays density ratio adjusted wind speed thresholds versus the density ratio adjusted winds in atmospheres at different times in the Moon's history.  Winds exceed required thresholds for Kelvin-Helmholtz wind wave generation in all cases and through to plagioclase floatation over nearly the entire physical extent of the atmospheres.  Capillary-gravity wave generation does not appear to be likely but even that mechanism may have been activated if some of the conservative assumptions regarding the atmospheres are relaxed (such as super-liquidus LMO temperatures and inclusion of latent heat of crystallization).  Even in cases with different orbital migration history of the Moon, we found that there were still wavelength bands of waves that would be generated.

Given the atmospheric profiles of horizontal wind velocity, we can also assess how quickly any floats would be advected from the Earthside to the far side.  A simple means of estimating the advection velocities is by determining the drag force of the wind on 'rockbergs' by using the drag equation.  For a hypothetical rockberg (using a drag coefficient of 0.002 from \cite{SMITH1983241}), the wind drag force is $\sim$ 0.002 * $\rho$ * $v^{2}$.  The variable $\rho$ is the density of the atmosphere while $v$ is the horizontal friction velocity.  Considering a dynamic viscosity for the magma given the kinematic viscosities listed before, the advection time for a 1 m rockberg from the sub-Earth point to the anti-Earth point is $\sim$ 15, 75 and 450 Earth years respectively, for the three times listed considered in the models.  Thus, winds were only able to drive floats completely to the far side if those floats formed greater than $\sim$400 years prior to when the bulk of plagioclase crystallization occurred. 

\subsection{Magma Ocean Cooling and Delayed Solidification}
\label{CoolSolid}

\begin{figure}[b!]
 \includegraphics[width=\textwidth]{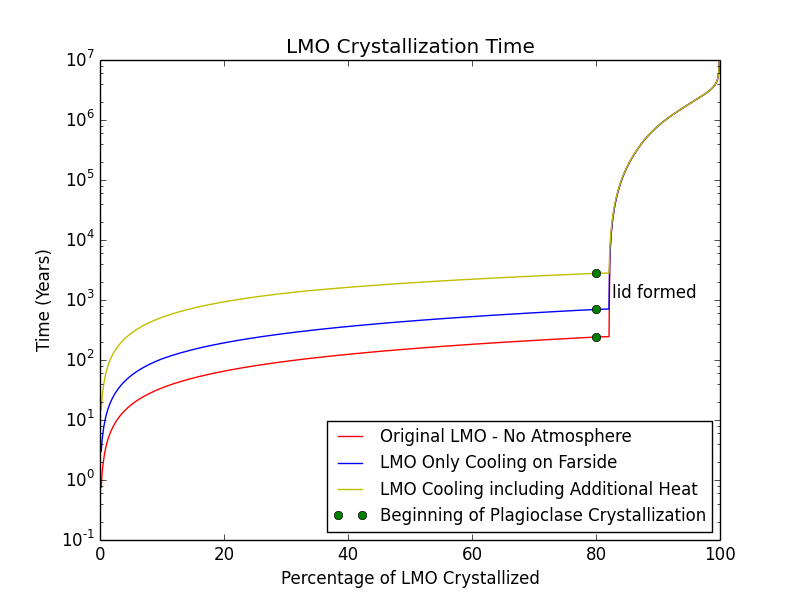}
\caption{Lunar Magma Ocean crystallization and plagioclase lid formation timescales for different cooling models.  Details regarding the different cooling models are given in section \ref{CoolSolid}.  While time to plagioclase lid formation may vary by an order of magnitude depending on the cooling model, times for total LMO solidification barely varies. }
\label{fig:cooling}
\end{figure}

The asymmetric radiative environment the surface of the post-collision  Moon was subject to has led some previous studies to suggest magma ocean cooling and solidification may have been delayed \cite{WASSON1980752, JGRE:JGRE1420, 2014ApJ...788L..42R}. For the effect of the modeled radiative environment on plagioclase solidification and lid formation in the magma ocean, we considered three different scenarios.  The effect of the three different scenarios on lid formation and total solidification is given in figure \ref{fig:cooling}.  The first is the original cooling model, described in \cite{2011E&PSL.304..326E}.  In this model the lid-less Moon cools radiatively to a vacuum over its entire surface.  Once the lid forms, it cools through conduction at the same rate over the entire surface.  The second model only allows the LMO to cool through net heat loss given by spatially summing the marginal outgoing flux during solidification.  This marginal outgoing flux is calculated by determining the incoming flux from external radiative sources and only allowing cooling by the excess radiative flux that is outgoing.  In this scenario there is no assumption that Moon must radiate any of the incoming flux (even in regions where incoming flux is greater than outgoing flux - which is true over a small portion of the moon immediately after formation). The Moon cools efficiently to a vacuum over the atmosphere-less region on the farside.  The Moon also cools somewhat efficiently over portions of the Earthside with lower surface temperatures.  However, because this scenario is dominated by farside cooling, we label it as 'Only Cooling on Farside' in figure \ref{fig:cooling}.  This model also switches to cooling by conduction after lid formation, but retains the marginal summing of fluxes to determine outgoing flux. In the final scenario, we again limit total cooling by an energy balance but now also require the moon to radiate away additional incoming flux.  This represents the other boundary condition on cooling as in reality most of the incoming flux is assumed to vaporize the magma to form the atmosphere.  While this heat will be deposited to some extent as the atmosphere is advected to cooler regions, the atmosphere itself will radiate to space (and will also lose energy through escape).  

\begin{figure}[b!]
 \includegraphics[width=\textwidth]{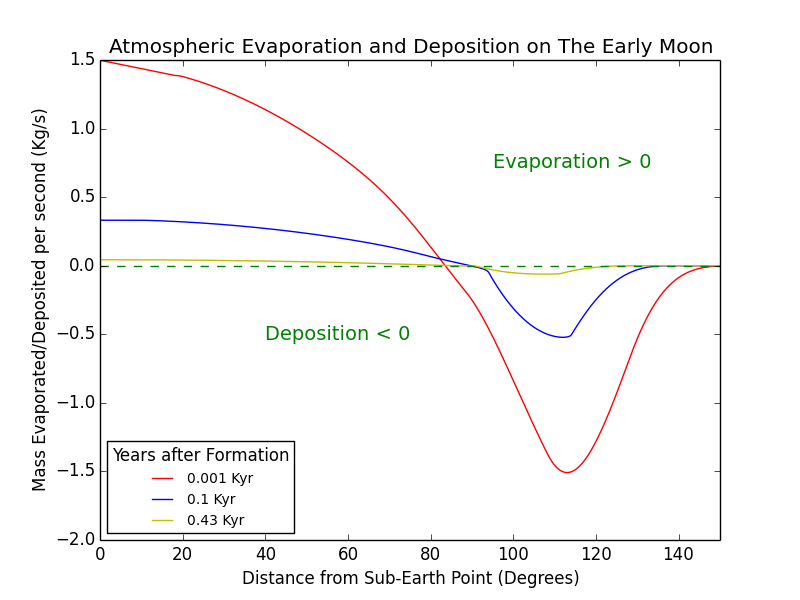}
\caption{Evaporation (positive values) and Deposition (negative values) profiles for the Early Moon.  The times correspond to orbital separations of 3, 4 and 5 Earth Radii.  Evaporation occurs over the majority of the Earth-facing hemisphere while deposition occurs in a relatively narrow band near the terminator region (which extends beyond 90 degrees).  There are slight changes in line shape near where the model uses an extrapolation to find a transonic solution. }
\label{fig:resurfacing}
\end{figure}

In the original cooling scenario, lid formation occurs at $\sim$250 years with a timescale of about $\sim$6 years from the beginning of plagioclase solidification to lid formation.  The farside cooling scenario extends those values by more than a factor of 2 (actually slightly less than a factor of 2 for time from plagioclase solidification to lid formation), while the final scenario extends time to lid formation by about about an order of magnitude and lid formation time by a little than a factor of 2.  In all scenarios, the time to complete solidification of the LMO does not change significantly given the strong dependence of the external incoming flux on orbital separation and Earth radiating temperature (which drops quickly to near present day values by about 10,000 years - see \citet{Zahnle201574,Zahnle2007}).  For the purposes of this study we assume cooling occurred and plagioclase floatation occurred on a timescale most similar to the middle scenario.  Even if the longest cooling scenario is assumed, the Moon would still have possessed a thin atmosphere to the floatation period since top of the LMO temperatures would also be slower to cool.

\subsection{Moderate Volatile Heterogeneities and Depletion through Escape}
\label{HetDep}

Finally, the last potential effects of the radiative environment on the early Moon may have been transport of moderate volatiles that may result in abundance heterogeneities and the potential that atmospheric loss could have contributed to the observed depletion of moderate volatiles.  

While atmospheric temperatures were likely too low to have influenced evaporation and deposition of more refractory materials, the moderate volatiles that dominated atmospheric composition were likely transported until, and likely even slightly after, lid formation.  The spatially decreasing temperatures and pressures of the atmosphere as it moved away from the sub-Earth point drove strong winds that transported energy and particles to the far side.  There existed distinct zones of evaporation and deposition that are given in figure \ref{fig:resurfacing}.  While the evaporation and deposition rates are unsurprisingly largest during earlier periods when the atmospheres were thickest, it is questionable whether they produced long lived spatial heterogeneities in magma composition.  On one hand, at least initially after formation, significant amounts of the dominant atmospheric constituents were vaporized and transported.  Since temperatures immediately declined to values too low to vaporize SiO (with evaporation rates falling closer to the 0.1 Kyr values), the dominant constituent vaporized after the intial $\sim$ 1 yr was sodium.  Using the values given in figure \ref{fig:resurfacing}, a significant fraction (up to potentially values greater than 1) of the total sodium content of the LMO would have been vaporized at some point in the first 100 yrs.  However, much of this transported and likely returned to sodium poor portions of the LMO due to what is expected to be rapid compositional planetary scale convection at these early times. Estimates for such convection range on the order of hours to days \cite{JGRE:JGRE3055}.  At those early times, the hot magma had such low viscosity that motion within the ocean was likely to disperse regions with elevated volatile abundances.  While regions of evaporation may have persistently exhibited depletion, such a conclusion would need to be understood by using a sophisticated model of magma ocean circulation.  

However, the period of time when plagioclase crystallization and flotation was occurring may have retained volatile abundance depletion and elevation.  For example, using just the evaporation and deposition rates given for the latest time in figure \ref{fig:resurfacing} (which was close to the time of lid formation), rockbergs that crystallized and buoyantly rose in different portions of the moon and were advected to the farside may have sodium abundance variations of several factors (and in particularly fortuitous starting points - more than an order of magnitude).  Additionally, the deposition profiles also indicate that at that time, regions towards the far side of the Moon beyond the terminator are likely to exhibit enhanced abundances of Na (and likely other less abundant volatiles ignored by our model, such as Potassium and Chlorine).  Whether these abundance variations and excesses would be able to withstand a long future of space weathering and rotational reorientation and still be present in lunar samples today, is an open question.

The potential of moderate volatile escape is also a pertinent question, given that the dominant constituents of these early atmospheres are the same elements that appear to be depleted relative to Earth abundances.  Scale heights near the sub-Earth point of the moon (where evaporation was greatest) may have approached 50-40 times the present day Earth's scale height for the times examined.  Thus thermal escape may have occurred more efficiently on the early Moon during the short period of time that the Moon was lidless.  Calculating thermal escape during this time is a poorly constrained problem, given the dependence of exobase height and escape rate on the vertical temperature profile of an atmosphere.  While our atmosphere model is purposely single component throughout most of the time examined, even trace amounts of other constituents may influence and control vertical structure and escape.  However, in order to put some bounds on escape rates, we estimate potential Jean escape \cite{JeansBook} using some simplifying assumptions.  Assuming an exobase temperature equivalent to the surface atmosphere temperature (we also tested temperatures 70 and 90 percent of this value with minimal change in escape estimates), we can calculate a lower bound on escape rate at the exobase height for atmosphere.  The integrated escape over the first 100 years (when escape was greatest) is 4-5 magnitudes smaller than the total sodium content of a primordial magma ocean.  However, this lower bound is unrealistic because the exobase height in this scenario is significantly inwards of the Roche limit for the Moon (even at 4 Earth radii separation) at the sub-Earth point.  As a more accurate estimate, we estimate the Jeans escape by taking the exobase height at approximately the sub-Earth point Roche limit as it varies over the first 100 years.  In this case, using a weighted exobase height (and consequently escape velocity) and atmospheric density over time, we find that atmospheric escape may have depleted anywhere from 5-20$\%$ of the initial sodium content of the LMO.  The exact value depends on the choice of the averaged exobase height.  However, this estimate also is an inaccurate measure of the likely escape.  Taking the exobase at this height ignores that the atmosphere would be collisional at such a height - which will act to surpress escape.  However, simple expressions for escape velocity and consequently atmospheric escape only consider the gravity of the primary, while in this case the gravity from the secondary is likely to make escape more efficient.

Additionally, recent research \cite{Airapetian2016} suggests that the Earth-Moon system existed in a far more violent space weather environment with a high frequency of paleo-solar eruptive events.  Given that a post-collision Earth and Moon were unlikely to possess strong coherent dynamos and consequently protective global magnetospheres, frequencies of $>$1 superflare generated incident CME per day may have served as an additional atmospheric loss mechanism.  Given that the early Moon's atmosphere was likely composed of volatiles vaporized from some layer of magma near the surface, even moderate atmospheric loss may have contributed to the observed depletion of moderate volatiles.  Such a process may not be the sole source of moderate volatile depletion given other mechanisms that may have also operated \cite{Canup2015}, but is certainly worth examining as a potential depletion pathway.  

\subsection{Alternate Moon Formation/Migration Models and Future Work}
\label{Alternate}

Our study relies on a number of assumptions, typically derived from the prevailing literature, that control the physical environment of the early Moon, and consequently our conclusions.  However, in some cases it is illustrative to consider alternative scenarios and their effects on the results of the study.  One of the primary assumptions we made was regarding the orbital history of the Moon.  Alternative migration histories may have led to slower or faster orbital migration of the Moon, particularly due to the influence of tidal effects.  Most cases in the literature suggest the Moon's orbital distance likely increased more slowly than we assume \cite{Zahnle201574, Chen2016132}. In the case of slower migration away from the Earth than we assumed, the early atmosphere is likely to have been even thicker and dominated by longer lived strong Earth to farside winds given the increased flux from the hot Earth.  There are fewer models that examine an orbital migration faster than the one chosen for this study (significantly greater than 6 Earth Radii by 2-3,000 yrs or earlier - in fact most don't expect near that separation until 10,000 years).  The are several model runs in \cite{Chen2016132} which examine very low values of Q for the Earth and fairly large tidal time lags.  However, the authors specifically reject them in the paper as not realistic. A model from \cite{0004-637X-760-1-83} suggests the moon progenitor had only mostly developed at a distance of about 4.25 Earth radii, 1 day after collision.  Even in this model, however, the outer bound at times of interest for this study was inside 6 Earth radii.  Model runs we ran based on this migration history still lead to magma ocean wave generation (though waves generated are in a narrower range of wavelengths).

While the assumption of a Moon which tidally locks to the Earth within 100 days is the prevailing one in most literature, it is also worth considering a case where Moon takes much longer to tidally lock (such as cases where the Earth had a high obliquity prior to the Moon forming collision). In this case the Moon would still experience the radiative inputs as delineated in our model, but would not have a permanently locked side that would result in an Earth-side/farside asymmetry in flux from the Earth.  In such a scenario, the Moon would likely have a global atmosphere still characterized by high temperatures and strong winds.  Given the similarity to the radiative environment of close-in exoplanets, the Moon would still have likely had very strong winds that may have generated waves in the magma ocean.  These waves are likely to have been unidirectional in the direction of the Moon’s rotation.  Plagioclase float motion in this scenario can then best be thought of by analogy to iceberg motion in the ocean - since ocean wave and wind drag are both proportional to the surface area of icebergs \cite{SMITH1983241}, different sized icebergs are advected at different rates.  A similar process would have occurred for the plagioclase ‘rockergs’, and assuming even a very low ‘sticking’ parameter, these different size floats may have eventually collected to form a large rockberg continent that may have been the seed for the lunar highlands.

An even more fundamental assumption of this study is that the Moon was formed from a collision between the Earth and a large planet-sized body.  The quick accretion of the ejected material deposits so much energy onto the Moon in such a short time that a deep global magma ocean is believed to have been formed.  However, alternative Moon formation models have also been proposed \cite{Rufu2017}. In certain cases where accretion of the Moon occurs on a longer timescale at a distance farther from the Earth or where there are multiple smaller impactors to the Earth, our study suggests observations which may differentiate between the scenarios.  In particular, for formation scenarios with a cooler post-collision Earth or a Moon that accreted at greater orbital separation, early lunar atmospheres would be much thinner and likely would not have transported such significant amounts of moderate volatile material.  Consequently, evidence of far side gradients and elevated abundances of moderate volatiles may lend credence to the canonical formation model. 

\section{Conclusion}
\label{sec4}

Under the canonical view of Moon formation, we find that it is likely that radiative environment the Moon was subject to resulted in a short lived metal atmosphere on the body that existed prior to plagioclase lid formation.  Such an atmosphere was likely characterized by pressures in the $10^{4}-10^{2}$ range, sub-Earth temperatures greater than 2000K, supersonic winds pointed to the far side and a metal composition.  While this atmosphere was unlikely to have greatly influenced cooling of the LMO, it did likely produce surface waves in the LMO due to the high winds (though wind drag driven plagioclase float advection appears unlikely).  Additionally, it appears likely that such an atmosphere may have transported and lost sodium and other moderate volatiles at significant enough rates to potentially produce heterogenties and depletion in the abundance of such elements.  Evidence of such heterogenities and depletion may help to constrain details regarding the formation process of the Moon. Such an atmosphere may also be a cousin to atmospheres expected on some of the most close-in and heated rocky exoplanets. 

\section{Supplementary Material}

\section{Atmosphere Model Details}
\label{sec1}

Our atmospheric model is nearly the same that was developed for Io in \citet{1985Icar...64..375I} and that applied to hot super-earths in \citet{2011ApJ...743L..36C}. It solves for the atmospheric pressure, temperature and velocity (at the base of the atmosphere) as a function of the angular distance $\theta$ away from the substellar point.  The model solves conservation equations for mass, momentum and energy:

\begin{equation} \dfrac{1}{rg \sin\theta}\frac{d }{d \theta}(VP\sin\theta) = mE \end{equation}

\begin{equation} \dfrac{1}{rg \sin\theta}  \frac{d } {d \theta} [ (V^2 + \beta C_{p} T)P\sin \theta ] = \dfrac{1}{rg \tan\theta}\beta C_{p}TP + \tau \end{equation}

\begin{equation} \dfrac{1}{rg \sin\theta}  \frac{d } {d \theta} [ (\dfrac{V^2}{2} + C_{p} T)VP\sin \theta ] = Q \end{equation}

Details regarding the formulation of these equations are available in \citet{1985Icar...64..375I} but a general description of the above formulas is that the left side of these equations are the vertically integrated divergence of horizontal fluxes of mass, momentum and energy while the right side represents the mass, energy and momentum transport between the atmosphere and ground.  The radius of the planet in meters is given by $r$ and $g$ is the surface gravity, which is simply $g =  GM/r^2$ with M being the mass of the planet and G being the gravitational constant.

$\beta$ is a thermodynamic parameter given by $\beta= (R/(R+C_{p}))$ where $C_{p}$ is the specific heat at constant pressure and $R = k_{b}/m$.  The values for the mass per atom, m, as well as $C_{p}$ are taken for the expected dominant constituent in the atmosphere (which is sodium for most times, with the exception of the intial model at 0.001 Kyr where models for SiO and Na are run).  The choice for this single constituent atmospheric model is based upon expected vaporization pressures for a Bulk Silicate Earth magma given in figure 2 of \citet{2011ApJ...742L..19M}.  Vaporization pressures as a function of temperature were fitted in order to produce Clausius Clapyeron relations given by $P_{s}(T_{s}) = Ae^{-B/T_{s}}$ for each of the dominant constituents.  For equilibrium temperatures up to nearly 3500K, the vapor pressure of the dominant constituent is nearly an order of magnitude greater than the next most significant constituent.  This justifies to first order our assumption of a single species atmosphere and these vapor pressure curves can be used to extract the constants used in our vapor pressure equations.  

The quantities mE, $\tau$ and Q represent the mass, momentum and energy transport rate per unit area between the atmosphere and the ground over the vertically small surface boundary layer. Specific prescriptions for the momentum and energy transport are based on the assumption that those transport rates are linearly dependent on the surface flux quantities being transported - which is reflected through the linear dependence of the equations on $\rho_{s}$. Equations for the particle, momentum and energy transport rate are given by:

\begin{equation} E = \dfrac{(P_{s}-P)}{v_{s}\sqrt{2\pi}} \end{equation}

\begin{equation} \tau = -\rho_{s}\omega_{a}V \end{equation}

\begin{equation} Q = \rho_{s}\omega_{s}C_{p}T_{s} -\rho_{s}\omega_{a} (\dfrac{V^2}{2} + C_{p}T) \end{equation}

Here, the particle flux E is proportional to the difference between the surface vapor pressure $P_{s}$ and the atmospheric pressure P.  The local speed of sound is $v_{s}$ which is given by $v_{s} =  (k T_{s}/m)^{1/2}$.  The surface boundary layer density $\rho_{s}$ is given by $\rho_{s}=(mP_{s})/(k_{b}T_{s})$. Momentum and energy transport rates are mediated by transfer coefficients $\omega_{a}$ and $\omega_{s}$.  Since momentum and heat transfer in the surface boundary occur by both advection by a mean flow normal to the surface and by eddy driven turbulence, theses coefficients are parameterized in terms of two different velocities representing each process.  The mean flow velocity $V_{e}$ is given by $V_{e} = mE/\rho_{s}$ while the eddy velocity $V_{d}$ is given by $V_{d} = V^2_*/V_{0}$ where $V^2_*$ is the frictional velocity.  For the turbulent flow observed in our models the frictional velocity is solved for iteratively using the equation $V_{n+1} = 2.5 V_{*} log[(9.0 z V_{*} \rho)/\eta] $ (for laminar flow one would find $V_{*} = [2 \eta V_{0}/\rho H]^{1/2}$).  The transfer coefficients are then written in terms of simple functions of $V_{e}$ and $V_{d}$ that model the behavior of the transfer coefficients in known limits of $V_{e}>>V_{d}$ and $V_{d}>>V_{e}$ - the exact description of these formulations are given in \citet{1985Icar...64..375I}. The dynamic viscosity $\eta$ is given by Sutherland's formula:

\begin{equation} \eta = \eta_{0}(T_{s}/T_{0})^{3/2}[(T_{0} + C)/(T_{s} + C)]  \end{equation}

with $\eta_{0}$ equal to $1.8 X 10^{-5}$ kg $m^{-1} s^{-1}$, $T_{0} = 291K$ and C = 120K (it is important to note that Sutherland's formula is only valid to about 555K, but our simulations show that our results are not highly sensitive to variations in $\eta$).

Finally, in terms of solving for P, V and T at a new location as one advances from the substellar point, one must first calculate the right hand side of the main conservation equations (2.1 - 2.3) using previous values of P, V and T at $\theta$.  Upon moving to the next value of $\theta$, those values should be augmented by the fluxes that are calculated for the terms in parenthesis on the left hand side.  This will then yield the following system of equations where the fluxes are represented by the f values:

\begin{equation} V_{0}P_{} = f_{1} \end{equation}

\begin{equation}(V_{0}^2 + \beta C_{p} T_)P_{0} = f_{2} \end{equation}

\begin{equation}(V_{0}^2/2 + \beta C_{p} T_{0})V_{0}P_{0} = f_{3} \end{equation}

One can then solve for the velocity at the new value of $\theta$ using the following equation:

\begin{equation} V_{0} = \dfrac{f_{2} \pm (f^2_2 - 2\beta(2-\beta)f_{1}f_{3})^{1/2}}{f_{1}(2-\beta)}  \end{equation}

with the Mach value given by $M = V_{0}[(1-\beta)(\beta C_{p}T_{0})]^{1/2}$.  As the solution is advanced, these system of equations and the model developed here should yield a profile of P, V, T, E, $\tau$ and Q as function of $\theta$.

\subsection{Model Assumption and the Validity of the Atmosphere Model}
\label{validity}

The uncertainty regarding the exact properties of the Moon forming collision and our limited understanding of atmospheric properties in such extreme irradiation environments justify the use of a low dimensional model that includes only the anticipated dominant physical principles for this project.  There are a number of second order factors which could be considered a posteriori, and these may too have an effect on the structure and dynamics of these atmospheres.  These second order effects (including some in the preceding section) were examined and were not included if their magnitude was not significant enough to affect the atmosphere or if their parameter space was too poorly constrained. The effects that were included are explicitly described in the paper - for example, accurate characterization of radiation from Earth that attempted to incorporate the angular size of the Earth as it appeared in the Moon's sky.  Additionally, the validity of atmosphere model as a reasonable approximation of the state of the atmosphere is based on a set of assumptions detailed in \cite{1985Icar...64..375I}.  Those assumptions and the corresponding tests that were conducted and passed are the following:

\begin{itemize}

    \item The atmosphere exhibits conservation of mass, momentum and energy.   There is also an additional requirement of surface conservation of particle flux during sublimation and resurfacing processes.  This is implemented by ensuring that the extrapolation chosen past the transonic point results in conservation of particle flux.  Atmospheric variables are scaled by a factor (which in all cases is very close to 1) in order to ensure this.
    
    \item The surface temperature of the Moon, which controls melting and vaporization processes, can be obtained by radiative balance estimates.  This is tested by examining the relative contribution of the latent heat flux of \textit{vaporization}, which is a small fraction of the radiative balance (typically $\sim$ 10$\%$ near the sub-Earth point).  For example, latent heat contributions would be less than $\sim$ 8$\%$ compared to radiative sources at the sub-Earth point at 0.1 Kyr.  This does not include latent heat flux of \textit{crystallization} of the LMO.

    \item The atmosphere behaves as a continuous fluid and is hydrostatically bound to the planet.  The mean free path of the particles must be smaller than the local pressure scale height - a necessary condition for the continuous fluid assumption.  We find that this criteria holds along the bulk of the flow and we stop model when this condition starts to break down (where the mean free path begins to exceed 10\% of the local scale height).  The local scale height must be small compared to the radius of the planet - while the scale height is relatively large over a small angular portion of the body very early after Moon formation, it is still considerably smaller than the radius in all cases.  Finally, kinetic energy of the atmospheric flow must be (and is) less than the binding energy of the planet.

    \item The fourth assumption is that the flow can be treated as turbulent and consequently that parameters such as the horizontal velocity and entropy per unit mass can be treated as constant with respect to the vertical axis.  The Reynolds number is used to verify that the flow is indeed, turbulent.

    \item The Moon is tidally locked to the primary radiation source (the Earth).  This is an assumption of the model that reflects the prevailing literature on the subject.  However, we discuss potential ramifications of relaxing this assumption in the main paper.
\end{itemize}

\section{LMO Model Details}
\label{sec2}

The Lunar Magma Ocean Solidification model used in our study is the model described in \cite{2011E&PSL.304..326E}.  This model integrates physical and chemical constrains of LMO solidification in order to determine timescales for crystallization/solidification, flotation crust thickness and post-overturn lunar mantle composition and structure.  It tracks the evolution of a convecting magma ocean with a depth of 1000 km.  

The model uses a lunar bulk silicate composition from \cite{1980LPSC...11.2043B} that is fractionally crystallized in steps that correspond to 1/1000 of the total LMO volume. Fractionation of magma ocean cumulates in
assemblages are determined
a priori.  Olivine is the sole mineral crystallized until solids have
filled the lunar interior to a pressure of 3.0 GPa.  After this, the crystallizing assemblage becomes 90$\%$ orthopyroxene and 10$\%$ olivine until nearly 80$\%$ of the LMO volume has crystallized.  This point is of particular relevance to the study, as after 80$\%$ crystallization, clinopyroxene and plagioclase crystallize.  The crystallization of the less dense (as compared to the co-existing liquid) plagioclase leads to the formation of the anorthite lid, which slows the cooling of the LMO.  Finally, the model crystallizes the remaining 15$\%$ or so of the LMO as a combination of minerals that include orthopyroxene and oxides in addition to the previous minerals.  The crystallization scheme and cooling rate of the LMO is fairly insensitive to the choice of a BSE composition (we tested both compositions with little difference in flotation and solidification timescales).  The primary reasoning behind the choice of the BSE composition was to remain agnostic about the source of loss of moderate volatiles given potential loss during the Moon's early atmosphere phase discussed in the main paper and potential future loss after LMO solidification. 

Cooling is modeled as a two stage process - through ocean convection and radiation from the surface until lid formation, and then using the transient heat equation once heat is radiated away through conduction following the formation of the lid.  In our study, we also consider a number of different cooling scenarios that incorporate energy balance from incoming radiation.  We calculate solidification and lid formation timescales using this cooling model using equations 4 and 5 from \cite{2011E&PSL.304..326E}.  The external temperatures are set using radiative balance to determine surface temperatures. In the LMO itself, material is assumed to retain its solidus temperature as crystallization proceeds.  The solidus for the evolving liquid decreases to the temperatures calculated using the MELTS program \cite{1995CoMP..119..197G} using equation 3 in \cite{2011E&PSL.304..326E}.




\bibliographystyle{elsarticle-harv}

\bibliography{lunarbib}

\end{document}